# RF system for the MICE demonstration of ionisation cooling

K. Ronald, C.G. Whyte, A.J. Dick and A.R. Young
SUPA, Department of Physics, University of Strathclyde Glasgow, UK and
Cockcroft Institute, Daresbury Laboratory, Warrington, UK

D. Li, A.J. DeMello, A.R. Lambert, T. Luo
LBNL, Berkeley, USA

T. Anderson, D. Bowring, A. Bross, A. Moretti, R. Pasquinelli, D. Peterson, M. Popovic, R. Schultz, J. Volk,
FNAL, Batavia, USA

Y. Torun, P. Hanlet, B. Freemire,
Department of Physics, IIT, Chicago, USA

A. Moss, K. Dumbell, A. Grant, C. White, S. Griffiths, T. Stanley[*], R. Anderson[*]
STFC, Daresbury/[*]Rutherford Appleton Laboratory, Warrington/[*]Chilton, UK

S. Alsari, K. Long, A. Kurup
Department of Physics, Imperial College, London, UK

D. Summers
Department of Physics and Astronomy, University of Mississippi, Mississippi USA

P.J. Smith,
Department of Physics and Astronomy, University of Sheffield, Sheffield, UK

*Abstract*—Muon accelerators offer an attractive option for a range of future particle physics experiments. They can enable high energy (TeV+) high energy lepton colliders whilst mitigating the difficulty of synchrotron losses, and can provide intense beams of neutrinos for fundamental physics experiments investigating the physics of flavor. The method of production of muon beams results in high beam emittance which must be reduced for efficient acceleration. Conventional emittance control schemes take too long, given the very short (2.2 microsecond) rest lifetime of the muon. Ionisation cooling offers a much faster approach to reducing particle emittance, and the international MICE collaboration aims to demonstrate this technique for the first time. This paper will present the MICE RF system and its role in the context of the overall experiment.

## I. INTRODUCTION

The Muon Ionisation Cooling Experiment (MICE) [1-3], Fig. 1, is being built to demonstrate that ionising interactions with low Z 'absorber' materials, specifically liquid $H_2$ or Lithium Hydride, followed by re-acceleration in RF cavities can significantly decrease the emittance of a muon beam in the momentum range of 140-240 MeV/c (simulations predict several percent, depending on initial emittance and energy spread). At MICE, a fraction of the proton beam in the ISIS synchrotron is periodically intercepted by a dynamically inserted target [4], producing a beam of pions which decay to provide the muon beam. The interaction of this beam with low Z absorbers in a strong magnetic field reduces all components of the particle momentum. RF cavities are required to restore the lost energy as ionization cooling is effective only over a limited range. Future muon accelerators would have a 'front end' with a repeating lattice of similar devices. Up to a factor of $10^5$ reduction in 6D emittance has been predicted in simulations of some linear 6D cooling systems [5].

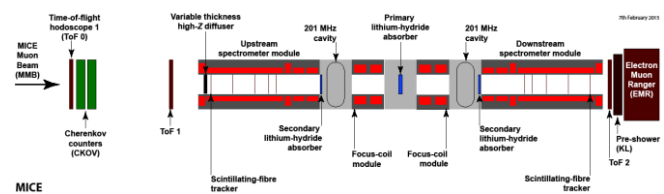

Fig. 1. Illustration of the key features of the MICE Demonstration of Ionisation Cooling

Strong magnetic fields are critical to effective ionization cooling, however immersing the RF cavities, operating at 8 MV/m or more, in this magnetic field gives rise to enhanced dark current and breakdown risk. MICE is simultaneously a physics experiment and a technology demonstration, showing that the RF accelerators can be integrated in the environment of strong magnetic fields. The paper will describe the key features of the RF system, and the principals of the MICE experiment.

## II. RF SYSTEM

### A. RF Cavities

The development of the RF cavities for MICE was a core R&D activity for the US Muon Accelerator Physics programme. Two cavities are required for MICE operating at gradients of up to 10.3 MV/m. The RF cavities are 201.25 MHz, room temperature, copper cavities (see Fig. 2) with large central beryllium windows for the beam [6]. Each cavity is fed from two 4" co-axial couplers mounted equatorially. A power of 1 MW to each cavity (500 kW on each coupler) gives rise to a gradient of some 8 MV/m. The particle density is very low, therefore there is no loading of the cavity fields by the beam and almost all the energy delivered to the cavity is dissipated in the walls of the cavity. The experiment is expected to operate at a repetition frequency of 1 Hz with pulses 1ms in duration. Each cavity will receive an average power of ~1 kW.

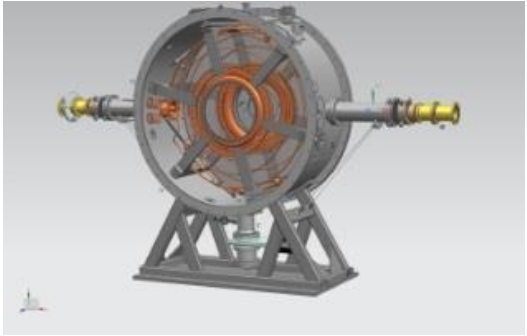

Fig. 2. Cut out view of single cavity module- the cavity body itself is transparent in this view. The central particle apertures and couplers are shown.

The RF cavities were electropolished and HP water rinsed (following SC cavity fabrication procedures) after spinning. In order to tune the cavity frequency, a system consisting of 6 stainless steel forks driven by an actuator system is used to vary the cavity body shape (see Figure 2).

A prototype system has been tested in the MuCool Test Area (MTA) at FNAL where gradients well in excess of those required were achieved (exceeding 14 MV/m) without sparking in the fringe field of a strong superconducting magnet [7]. The production cavity modules are nearing completion at LBL.

*B. RF Drive Systems*

To provide the input signal to the RF cavities an amplifier chain has been developed by the STFC Daresbury laboratory [8]. The system needs to provide pulses of 1 ms in duration at 1 Hz with a nominal output power of 2 MW. The system amplifies a low level (mW) drive signal using a 4 kW SSPA which drives a 250 kW tetrode valve amplifier (Photonis 4616) which in turn drives a 2 MW triode valve final stage amplifier (Thales 116). These amplifiers are supported by a custom developed modulator and protection system. The first amplifier chain has demonstrated the required performance and has been installed at MICE [9]. Many components required for a second amplifier are available. It is presently intended to use one amplifier which, with allowance for line loss and LLRF overhead, is expected to deliver 800 kW to each cavity for 7.3 MV/m gradient. The proposed installation is shown in Fig 3. If two amplifiers were to be installed this would provide a potential gradient of 10.3 MV/m. LLRF control systems based on the LLRF-4 system are being developed by the ISIS Injector RF group building on an approach developed by the LLRF group at Daresbury. This will control the phase and amplitude of the RF signal. The RF cavities will be kept on tune by a separate but closely linked system.

*C. RF Diagnostics*

MICE is designed to provide a relatively low flux rate such that individual particles are analysed. This is an essential feature to allow the required precision measurement of emittance. The particles are subsequently assembled into effective 'analysis bunches' to develop an understanding of how an ensemble beam would evolve. As the particles arrive asynchronous with the RF phase, it is important to associate each particle with the RF phase that it experiences at the cavity [10] to enable this analysis. This requires a measurement of the RF phase at the instant when the particles traverse the hodoscopes and then projecting the particle track to the cavity. RF waveforms from the tests at FNAL have been used to evaluate sub-sample measurements of the cavity phase when a particle transits the hodoscopes with a precision of <20 ps. The principles of this approach, methods to align the particle measurement and the RF measurement and the measurement of the RF phase by TDC will be presented.

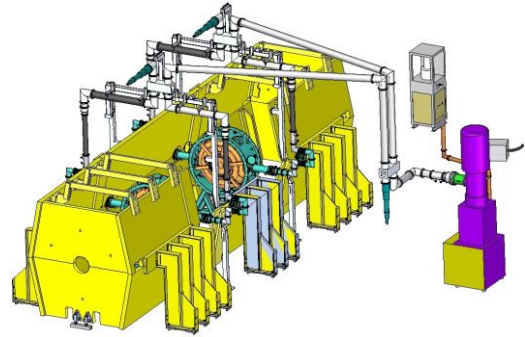

Fig. 3. Illustration of overall arrangement for single RF driver for 2 cavities. The yellow object represents the iron return yoke.

III. SUMMARY

The cavity tests for the MICE experiment have been completed successfully, whilst the essential first RF drive system has been proven. The production cavities are nearing completion and progress has been made on a second RF drive system. Work is continuing on the LLRF and controls systems and the RF diagnostics.

Sponsored by: DoE and NSF (USA); STFC (UK); EC FP7 (TIARA project, grant agreement no. 261905)